\documentstyle[aas2pp4,psfig,rotate]{article}
 
\newcommand{\gtsim}{\ {\raise-0.5ex\hbox{$\buildrel>\over\sim$}}\ }
\newcommand{\ltsim}{\ {\raise-0.5ex\hbox{$\buildrel<\over\sim$}}\ }
\def\Msun{\hbox{$\thinspace M_{\odot}$}}

\journalid{}{}
\articleid{11}{14}  

\begin{document}

\title{Some Constraints on The Formation of Globular Clusters}

\author{Keith M. Ashman}
\affil{Department of Physics and Astronomy, University of Kansas,
	Lawrence, KS 66045; and Department of Physics, Baker University,
	Baldwin, KS 66006 \\ e-mail: ashman@kuspy.phsx.ukans.edu}

\author{Stephen E. Zepf}
\affil{Department of Astronomy, Yale University, New Haven, CT 06520
\\ e-mail: zepf@astro.yale.edu}

\begin{abstract}
We explore the constraints on globular cluster formation provided
by the observed conditions in starbursts where globulars are currently
forming, and by the observed properties of young
and old globular clusters.  We note that the 
pressure in the ISM of starbursts and mergers implies that
molecular clouds in these environments have radii similar to those
of globular clusters.  Such molecular clouds are therefore viable
precursors to globular clusters if the star formation efficiency in the
clouds is high. A high star formation efficiency may be a consequence
of the high density and associated high binding energy
and short dynamical timescale of molecular clouds in such environments.
We also note that the apparent lack of a mass-radius relationship in 
young and old globular cluster systems places important constraints
on globular cluster formation models. This is because molecular clouds
are observed to follow a virial scaling relation between mass and radius. 
We suggest that a variable
star formation efficiency may weaken or eliminate the mass-radius
relation of molecular clouds as they fragment to form globular
clusters.  We attribute the absence of young 
globular clusters in the disks of ordinary galaxies such
as the Milky Way to the relatively low ambient pressures
in such systems. 

\end{abstract}

\keywords{globular clusters: general --- stars: formation}

\section{Introduction}

Globular clusters are found around nearly all galaxies, from low-luminosity
dwarf spheroidals to the most massive elliptical galaxies. However,
globular clusters are not currently forming in any significant
numbers in the disks of ordinary
galaxies like the Milky Way.
Models of globular cluster formation must therefore
address two related questions. First, what physical processes are
responsible for the formation of dense, massive star clusters in a broad
range of environments? Second, why is globular cluster formation
common in starbursts, but rare 
in the star formation regions of disk galaxies? 

The absence of young globular clusters in the Milky Way has compounded
the difficulty in understanding globular cluster formation. Indeed,
until relatively recently, globular clusters were often viewed as being
exclusively old objects. As a result, most globular cluster formation
models have been divorced from direct observation of 
star cluster formation and have instead been based on ideas concerning
the physical conditions at the pregalactic and protogalactic epochs.
For example, Peebles and Dicke (1968) suggested that  
globular clusters are the first bound structures
to form in the universe and therefore predate galaxies.
 Subsequent hierarchical clustering ensures that at least
some of the original population of globulars end up in the halos of galaxies.
Such ``primary'' formation models have fallen from favor for a variety of
reasons. These include the presence of a thick disk population of metal-rich
globular clusters in the Milky Way (Zinn 1985 and references therein),
correlations between the location of globular clusters in their parent
galaxies and globular cluster properties such as metallicity
(e.g.\ Harris 1991),
and the absence of dark matter halos around globular clusters (Moore
1996) that would
be expected if these objects had a cosmological origin (see Ashman and Zepf
1998 for a detailed discussion).

The above considerations moved the focus of globular cluster formation models
to the protogalactic epoch.
One influential model is due to Fall \& Rees (1985) who
suggested that protoglobular clouds form through thermal instability
in protogalaxies.
Fall \& Rees (1985) argued that such clouds would cool to around
$10^4$~K. At this temperature, cooling in a primordial (very low-metallicity)
plasma becomes inefficient under equilibrium conditions, in which case
the clouds ``hang up'' at $10^4$~K. The Jeans mass of such clouds is
around $10^6 M_{\odot}$ which is in the same ballpark as the
median mass of Milky Way globular clusters. Similar scenarios were
proposed by Gunn (1980), McCrea (1982) and Murray \& Lin (1992).

Further investigation of this problem has shown
that it is extremely difficult to maintain protoglobular
clouds at a constant temperature for a sufficient time that a 
characteristic mass is imprinted on an initial cloud mass spectrum.
Even in a primordial plasma non-equilibrium cooling can generate
significant amounts of H$_2$ which provides additional cooling
(Ashman 1990 and references therein).
Moreover, the high metallicity of many
globular clusters, such as the metal-rich population found in 
elliptical galaxies and the thick disk clusters of the Milky Way, suggests
that their precursor clouds were pre-enriched with metals. The presence
of such metals allows clouds to cool efficiently well below $10^4$~K. The
dependence of cooling rate on metallicity also means that Fall-Rees type
models predict that the mass of globular
clusters should decrease with increasing globular cluster metallicity.
Such a trend is ruled out observationally (e.g.\ Djorgovski \&
Meylan 1994).  Further, the physical conditions required by
thermal instability models do not exist in environments where globular 
clusters currently form, nor would they have ever existed in the lowest
mass galaxies with globular cluster systems (Ashman \& Zepf 1998).

Ideas concerning globular cluster formation have been greatly advanced 
by HST observations of young star clusters
in galaxy starbursts and mergers (e.g., Schweizer 1998; Ashman \& Zepf 1998).
Specifically, HST images
reveal that widespread formation of compact, bright, blue
star clusters is a characteristic feature of vigorous starbursts.
These young star clusters have sizes and estimated masses
like those of old Galactic globular clusters. 
The one aspect of the young globular cluster systems that is
different than that of old systems is their inferred mass
functions. Old globular cluster systems are well-known to have a 
log-normal mass function, while the mass function inferred for
young systems is consistent with a single power law over a
range that includes the mass scale at which the old cluster
systems begin to decline in numbers (Whitmore et al.\ 1999;
Zhang \& Fall 1999; Zepf et al.\ 1999; 
Carlson et al.\ 1998; Miller et al.\ 1997). However, since destructive
processes are expected to preferentially remove low-mass clusters from
the young cluster population, these observations are consistent with
the idea that systems of young clusters in starbursts will evolve to
resemble the traditional old systems of globular clusters (e.g. Fall \&
Rees 1977; Murali \& Weinberg 1996; Gnedin \& Ostriker 1997; Vesperini
1997).

It has been pointed out by several authors (e.g., Elmegreen \& Efremov 1997)
that the mass function
slope of Giant Molecular Clouds (GMCs) in the Milky Way and
other galaxies is similar
to that of young globular clusters in mergers. Given that GMCs are 
known to be the sites of open cluster formation, it is reasonable
to suppose that similar molecular clouds are the precursors of
young globular clusters. The similarity in mass functions adds weight to
this idea. However, the star formation efficiency in GMCs in galaxies
like the Milky Way is only around 0.2\%. For the typical masses of such
GMCs, this efficiency produces open clusters with masses up to
$10^3\Msun$ or so. Thus in order to produce clusters spanning the full mass
range of observed globular clusters, we either
require more massive GMCs (Harris \& Pudritz 1994) or we require
a much higher star formation efficiency. 

In this paper, we present an analysis of the globular cluster formation
process based on observations of the physical conditions in which globular 
clusters currently form, as well as 
observations of young globular clusters themselves.
Specifically, we consider the conditions in the ISM of merger-induced
starbursts and the properties of young globular clusters in this environment. 
Our results indicate that GMCs in starbursts
are viable globular cluster progenitors, but that the mass-radius
relation of young globular clusters places stringent constraints on
the details of the globular cluster formation process.
For GMCs with a power-law mass distribution like that in the Milky Way,
the high pressure that is characteristic of the ISM in starbursts 
compresses GMCs so
that they have the appropriate radii to become globular clusters, while
the high star formation efficiency found in starbursts produces much more
massive, tightly bound clusters than those formed in GMCs in the
Milky Way. However, the lack of a significant mass-radius
relation for young globular clusters (Zepf et al 1999),
[also found in systems of open clusters (e.g.\ Testi, Palla
\& Nota 1991; van den Bergh 1991) and old globular clusters 
(van den Bergh et al 1991; Djorgovski \& Meylan 1994; Ashman
\& Zepf 1998)] is surprising, since any system of
virialized gas clouds is expected to have a strong mass-radius relation. 
Thus a critical constraint on globular cluster formation models is
that the mass-radius relation of precursor clouds must be eliminated.
We show in this paper
that a star formation efficiency that varies with the binding
energy of the precursor clouds can wipe out a virial mass-radius relation.

In Section~2 we summarize the observational properties of systems of GMCs
and globular clusters. We show in Section~3 that GMCs in the
high-pressure ISM of starbursts are promising globular cluster precursors
and discuss physical mechanisms that may lead to a high star formation
efficiency in such clouds. Section~4 addresses the difference between
the mass-radius relation of GMCs and globular clusters and the possibility
that this difference may be the result of a star formation 
efficiency that varies with cloud 
binding energy. In Section~5 we present our conclusions.

\section{Observational Background}

In this section we describe the observed properties of GMCs and 
of young and old
globular clusters. These properties must be incorporated into any model of
globular cluster formation in which GMCs are globular cluster progenitors. 
Specifically, we discuss the mass and radius
distributions of GMCs and globular clusters and the mass-radius relations
followed by these objects. We review the evidence for the similarity at high
masses of the GMC and globular cluster mass distributions. 
In contrast, the mass-radius
relations followed by these objects differ.
We also compare the radius distributions of globular
clusters and GMCs. 
As we show in Section~4, these
observations place strong constraints on any scenario in which
globular clusters form from molecular clouds.

\subsection{The mass-radius relation of clouds and clusters}

A large body of observational evidence indicates that
GMCs in a range of environments follow a mass-radius relation that
is consistent with these objects being in virial equilibrium. 
Specifically, GMCs follow the scaling relation $r_c \propto M_c^{1/2}$
(Larson 1981), 
where $M_c$ is the cloud mass and $r_c$ is the cloud radius.
This has been established for GMCs in the Milky Way (see the summary
of observations
given by Harris \& Pudritz 1994), as well as M33, the LMC and
the SMC (Wilson \& Scoville 1990; Johansson 1991; Rubio et al.\ 1993).

In sharp contrast to GMCs,         
globular clusters appear to have a weak or non-existent correlation
between radius and mass. This is relatively well-established for
the Galactic globular cluster system (van den Bergh
et al 1991; Djorgovski \& Meylan 1994; Ashman \& Zepf 1998) and seems to
hold for the young globular
cluster system in the merger/starburst NGC~3256 (Zepf et al.\ 1999), where
there may be a shallow correlation.
A similar weak or absent correlation between mass and radius also
appears to hold for the young star clusters in the LMC (van den Bergh 1991),
and for young star clusters in the Galaxy (e.g.\ Testi,
Palla, \& Nota 1999).
In order to wipe out the mass-radius
relation of the parent GMCs, it is apparent that either the mass
or the radius (or both) of the final star clusters must differ from those
of the original clouds. This difference in the
mass-radius relations provides a challenge for any model in which globular
clusters form from GMCs in virial equilibrium, particularly given the
similarities in mass and radius distributions between GMCs and 
globular clusters discussed below. We return to this point
in Section~4 and discuss how the difference might arise.

\subsection{The mass and radius distributions of GMCs and globular clusters}

The mass distribution of GMCs is usually expressed
in the form:
$$
N(M_c)dM_c \propto M_c ^{-\beta} dM_c .
\eqno(2.1)
$$   
Published values of $\beta$ vary from study to study. 
Harris \& Pudritz (1994) summarize results from CO surveys in which the
mass distribution of Milky Way 
GMCs is derived from the observed distribution of cloud sizes and the
observed relation between cloud velocity dispersions and sizes
(essentially a mass-radius relation). The result is $\beta =1.63 \pm0.12$.
A similar relation is found for clouds in M33, and for the LMC and SMC.
Elmegreen \& Falgarone (1996) analyze
more recent results and obtain $\beta$ in the range 1.5--2. 
However, they also emphasize the problem of systematic
errors in mass measurement and suggest that without such errors the
value of $\beta$ may be close to 2 in a wide range of environments.

The mass function of clusters may be written in a similar fashion:
$$
N(M_*)dM_* \propto M_* ^{-\alpha} dM_* ,
\eqno(2.2)
$$
where, for young clusters, $\alpha \simeq 1.8$
(e.g.\ Whitmore et al.\ 1999; Zhang \& Fall 1999;
Zepf et al.\ 1999; Carlson et al.\ 1998; Miller et al.\ 1997).
The similarity between the mass function slope of GMCs and young globular
clusters has been noted by several authors (e.g., Elmegreen \& Efremov 1997)
and is an important motivation for our identification of GMCs as globular 
cluster progenitors.  This value is also
consistent with the high-mass end of the mass function of old globular
clusters (e.g., Harris \& Pudritz 1994),    
with the possible exception of the most massive clusters
(e.g.\ Burkert \& Smith 2000 and references therein).
The fact that GMCs follow a mass-radius relation means that their mass
and radius distributions are not independent quantities.
We can express the GMC radius distribution in a manner analogous to the mass
distributions given above:
$$
N(r_c)dr_c \propto r_c^{-\kappa} dr_c
\eqno(2.3)
$$
The studies summarized by Harris \& Pudritz (1994)
yield $\kappa \simeq 3.3 \pm0.3$. 

We can relate the mass
and radius distributions by noting that
$$
N(M_c)dM_c = N(r_c) \left( \frac { dr_c }{ dM_c } \right) dM_c
\eqno(2.4)
$$
Assuming a mass-radius 
relation of the form $M_c \propto r_c^x$, it follows that
$$
N(M_c)dM_c \propto M_c^{(2-\kappa-x)/x}dM_c
\eqno(2.5)
$$
Comparing equations (2.1) and (2.5) we find:
$$
\beta = \frac{ x + \kappa - 2 }{ x }
\eqno(2.6)
$$
or
$$
x = \frac{ \kappa - 2 }{ \beta - 1 }
\eqno(2.7)
$$
The observed values of $\kappa$ and $\beta$ yield $x \simeq 2.2$. Assuming
that the clouds are in pressure equilibrium, this is consistent
with the virial value of $x=2$ within the observational uncertainties.

The young and old globular clusters do {\it not} exhibit a significant
mass-radius relationship, therefore we can not relate their mass and radius
distributions as we did for GMCs. That is, $dr_*/dM_*=0$,
so there is no useful
analog to equation (2.4) for globular clusters. However, it is still interesting
to compare the radius distributions of globular clusters and GMCs.
As before, we express the radius distribution of clusters in the standard way:
$$
N(r_*)dr_* \propto r_*^{-\eta} dr_*
\eqno(2.8)
$$ 
For old globular clusters, $r_*$ is most usefully interpreted as
the half-light radius, $r_{1/2}$, since this radius has been shown to be
resilient to dynamical evolution effects
(e.g. Spitzer \& Thuan 1972; van den Bergh et al.\ 1991 and references
therein). However, a direct comparison between the radius distribution
of globular clusters and GMCs is difficult to interpret, even if 
$r_{1/2}$ is used in the comparison (see also Section 4 below). 
Part of the difficulty stems from the fact that the range of cluster
radii in young and old systems is less than the range in mass, so 
radius distributions are more uncertain than mass distributions. 
Further, the radii of globular 
clusters and GMCs are determined in different ways. Finally, 
the density profiles of GMCs are much shallower than those of globular
clusters, so significant evolution of the density profile, and thus
the cluster radius, is likely to occur during the cluster formation
process. We return to this point in Section~4.

With these caveats in mind, we find that
for clusters of large radii the radius distribution
of old Milky Way globulars and the young clusters in NGC~3256 can be fit
by a power law with $\eta \simeq 3.4$. In these calculations we used
the observational
data of Zepf et al (1999) for the NGC~3256 clusters and the McMaster catalog
(Harris 1996)
for Milky Way globular clusters.  This slope is consistent with that
of the radius distribution of GMCs quoted above. 
However, the similarity between GMC and globular cluster radius 
distributions {\it only} appears to apply to objects with large radii.
Ashman \& Zepf (1998) showed that the
old globular clusters of the Milky Way have a half-light radius distribution
that exhibits a distinct peak when plotted logarithmically.
There is currently only limited
information on the radii of young globular clusters, with the most complete
study being that of the NGC~3256 system (Zepf et al.\ 1999).  These young
clusters also exhibit a peak in the logarithmic radius distribution. 
A similar peak appears to be present in the young cluster system of the
Antennae (Whitmore et al.\ 1999). There is no evidence for such a peak
in the logarithmic radius distribution of GMCs down to the smallest scales
that have been observed (Elmegreen \& Falgarone 1996 and references therein).

In terms of understanding globular cluster formation,
the differences between the radius distributions of globular clusters and
GMCs may be more important than the similarities.
Recall that the mass distributions
of GMCs and {\it young globular clusters} are both well-approximated
by single power laws; it is only the old globular clusters that show
a turnover in their logarithmic mass distributions. This suggests that
the physical processes responsible for the mass distributions of old
globular clusters operate on long timescales. In contrast, differences
between the radius distributions of GMCs and globular clusters are
already apparent in young globular cluster systems, although the data
are currently limited. If these results are confirmed by subsequent 
observations, they will suggest that the deviation from a simple power
law of globular cluster radius distributions is produced by processes
occurring when the clusters are young or still forming.

\section{The Viability of Molecular Clouds in Mergers and Starbursts 
as Globular Cluster Precursors}

One additional piece of evidence that motivates
the scenario discussed in this paper is that
young globular clusters are found in mergers and starbursts,
but not in the disks of less active star-forming galaxies (with some
possible exceptions; see Larson \& Richtler 1999). This suggests 
that in order
to understand the physical conditions that give rise to globular cluster
formation, one should examine differences between the conditions in the
ISM of starbursts and that of ordinary disk galaxies. 
One notable difference is that the
ISM pressure in galaxy mergers is likely to be several orders of
magnitude higher than in the Milky Way. This is expected based on
the results of simulations (e.g.\ Mihos \& Hernquist 1996) and seen
in observations (e.g.\ Heckman, Lehnert \& Armus 1993;
Heckman, Armus \& Miley 1990). 
In this section we show that these higher
pressures lead to molecular clouds with properties appropriate for
globular cluster precursors.

We noted above that the mass function of
GMCs in the Milky Way and other galaxies has a slope which is similar
to that observed in young globular clusters in galaxy mergers and the
high-mass end of the mass spectrum of old globular clusters.
Harris \& Pudritz (1994) defined a ``median'' mass of Milky Way GMCs
such that half the mass in GMCs is in clouds with masses exceeding this
value. This median mass is dependent on the slope of the GMC mass function
and the assumed upper mass limit of clouds, but for characteristic
values such as those 
used by Harris \& Pudritz (1994) is around $3 \times 10^5 M_{\odot}$. 
This is around
2 to 4 times higher than the median mass of old Milky Way globulars. 
If such GMCs fragmented into bound star
clusters with a high star formation efficiency of around 25\% to 50\%,  
they would therefore produce clusters with a mass and mass spectrum
consistent with that of young globular cluster systems (and also of old
globular cluster systems if dynamical evolution removes low-mass
clusters).  However, the resulting star clusters would {\it not}
resemble globular clusters in terms of characteristic radii. The median
half-light radius of globular clusters in the Milky Way is 3.0~pc, whereas
the median radius of GMCs is around 20~pc.

Qualitatively, it is apparent that a higher external pressure will produce
more compact molecular clouds at a given mass. We can quantify this
by utilizing the Ebert-Bonner relations (Ebert 1955; Bonner
1956; see also Harris \& Pudritz 1994 and McLaughlin \& Pudritz 1996)
for self-gravitating, pressure-bounded isothermal spheres. These
may be written as:
$$
M_c = \frac{ 3.45 }{ \gamma ^{3/2} } \frac{ \sigma ^4 }{ (G^3P_s)^{1/2} }
\eqno(3.1)
$$
$$
r_c = \frac{ 0.69 }{ \gamma ^{1/2} } \frac{ \sigma ^2 }{ (GP_s)^{1/2} }
\eqno(3.2)
$$                        
where $M_c$ and $r_c$ refer to the mass and radius of the cloud as before,
$P_s$ is the surface pressure of the cloud, 
$\gamma$ is a factor of order unity which is dependent on the
nature of the equilibrium, $\sigma$ is the one-dimensional
velocity dispersion within the cloud,
and $G$ is the gravitational
constant. 
These equations lead to the scaling relation
$$
r_c \propto M_c^{1/2}P_s^{-1/4}
\eqno(3.3)
$$
This scaling can be simply understood by expressing the
virial theorem in terms of temperature
$T \propto M_c/r_c$, and noting that density
$\rho \propto M_c/r_c^3$ and (assuming that the clouds are
in pressure equilibrium) $P_s \propto \rho T$.

The Ebert-Bonnor relations are believed to provide a reasonable 
quantitative description of GMCs in the Milky Way 
(e.g., Elmegreen 1989). Therefore,
the scaling relation (3.3) allows us to calculate the equilibrium
properties of GMCs in the high pressure environment
of the ISM in a merger,
assuming that GMCs in such an environment are in virial equilibrium.
If we consider the median Milky Way GMC with a radius of 20~pc and increase
the surface pressure by 100
we find the radius of the cloud to be 6.3~pc. Given that
this is the total radius of the cloud, and that the median half-light radius
of old Milky Way globulars 
(and young globulars in mergers) is around 3.0~pc, it
is immediately apparent that our compressed GMC has properties consistent
with those of a protoglobular cloud. More generally, the above
equations imply that in the high-pressure ISM of starbursts
any pressure-supported, virialized molecular cloud with a mass
comparable to that of a globular cluster will also have a
radius characteristic of a globular
cluster. This is {\it not} the case for the much lower pressures
in the ISM of the Milky Way in which the GMCs are much less dense than
typical globular clusters.

It is important to note in this context that there is little direct
information about the nature of GMCs in starbursts. 
It is possible that in merger-driven starbursts the GMCs are simply
those from the ISM of the progenitor spirals. In this case, the above
calculation indicates that
such GMCs are viable globular cluster precursors, provided they reach
pressure equilibrium with the high-pressure starburst ISM {\it before}
they fragment to stars. It is also possible that some
or all of the GMCs in starbursts form within the high-pressure ISM.
In this case, the clouds form with radii like those of globular
clusters. Irrespective of these details,
the critical point is that
any viriliazed molecular cloud with a mass characteristic of a globular
cluster will also have a radius appropriate for a globular cluster,
{\it provided} the molecular cloud is in a high-pressure environment like
the ISM of starbursts.

\subsection{The Star Formation Efficiency in Compressed GMCs}

        A general requirement for the formation of {\it any} bound
star cluster is that the star formation efficiency must be {\it locally}
high: somewhere around 25\% to 50\%. 
If we consider a model in which protoglobular
clouds are compressed GMCs with masses like those in the Milky Way,
the fact that the median mass of GMCs is a factor of 2 to 4 greater than 
that of globular
clusters implies a {\it global} star formation efficiency around this value
is required to produce globular clusters.
We regard it as promising that the star formation efficiency required
by both arguments is the same.
It is also promising that the
starburst/merger environment in which GMCs might be compressed
is observed to have a higher efficiency of star formation overall. 
The physical connection between high pressure and high star formation efficiency
has been discussed by several authors (notably Jog \& Solomon 1992;
Jog \& Das 1996; Elmegreen and Efremov 1997).

The critical question is how
to achieve such a high star formation efficiency in a GMC. 
A notable feature of a compressed GMC is that its dynamical timescale of around
$10^6$ years is very short (see also Elmegreen \& Efremov 1997). This
is comparable to the timescale of various disruptive processes associated
with massive stars such as ionization fronts, stellar winds, and 
supernova explosions. These processes have the potential to terminate
star formation in GMCs and, if this results in a low star formation
efficiency, may lead to the resulting star cluster becoming unbound.
It is therefore plausible that within a 
compressed GMC the short dynamical timescale will allow
a large fraction of the original gas
mass to be converted into stars. In other words, clusters forming from
high density GMCs do so with a high star formation efficiency.
On the other hand, in lower density GMCs
such as those in the low-pressure ISM of the Milky Way, disruptive
processes associated with massive
stars may regulate or
terminate star formation before much fragmentation has occurred.

We can investigate the feasibility of this idea
by considering the scaling of
dynamical time of virialized clouds with a range of external pressures.
For clouds with
constant density as a function of radius, the dynamical 
time may be written as:
$$
t_d = \left( \frac{ 3 \pi }{ 32 G \rho } \right) ^{1/2}
    = \left( \frac{ \pi ^2 r_c^3 }{ 8 G M_c } \right) ^{1/2}
\eqno(3.4)
$$
We can now use the Ebert-Bonner relations to eliminate $r_c$.       
This yields:
$$
M_c \simeq 2 \times 10^3 \gamma ^{-3/2}
\left( \frac{ t_d }{ 10^6~{\rm yr} } \right) ^4
\left( \frac{ P_s }{ 10^5 {\rm k}~{\rm cm}^{-3}{\rm K} } \right)^{3/2}
\eqno(3.5)
$$
where $M_c$ is in units of solar masses.
$P_s$ is normalized to a value appropriate to
GMCs in the Milky Way, which is believed to be 
about a factor of five greater than the typical 
ISM pressure because of the HI envelopes surrounding
GMCs (Elmegreen 1989; see also Harris \& Pudritz 1994). 

We can use the above relation to estimate the maximum mass of Ebert-Bonner
clouds that, at a given surface pressure, have a dynamical time less than 
$10^6$~years.  For surface pressures
around the $10^5~k$~cm$^{-3}$K characteristic of GMCs in the Milky Way, the
maximum cloud mass satisfying this requirement is around
$10^3~M_{\odot}$. For surface pressures around $10^7-10^8~k$~cm$^{-3}$K
expected for clouds in the ISM of starbursts, 
this mass increases to around $10^7 M_{\odot}$. Although this is
clearly a simplified calculation, it illustrates an important trend.
Specifically, higher external pressures lead to an increase in the
maximum cloud mass for which the dynamical time of the cloud is 
less than timescales characteristic of disruptive stellar processes.  Despite
this simplification, however, the central idea that GMCs in a high-pressure 
environment are denser and therefore fragment more rapidly will still
apply in a more realistic treatment.

Finally, it is worth noting in this context
that the possibility of rapid fragmentation and star formation
is highly pertinent to one of the most striking observational characteristics
of globular clusters: the star-to-star homogeneity in the abundance of
iron-peak                                   
elements. If star formation occurs on a timescale comparable to or less than
the lifetime of massive stars, metals from supernovae will not end up
in low-mass globular clusters stars observed today. In this case, the only
effect of supernovae will be to purge young globular clusters of gas,
which is required anyway. This point has been made by Elmegreen and Efremov
(1997) in the context of their model of cluster formation.

\section{From Clouds to Clusters: A Variable Star Formation Efficiency}

The results summarized in Section~2 lead to the important conclusion that any
model of globular cluster formation (and cluster formation in general)
needs to explain the difference between the mass-radius 
relation of star clusters and their precursor clouds. In other words, an essential 
ingredient of any globular cluster formation model is a mechanism 
which modifies the virial mass-radius relation of the original clouds.

In this section we investigate whether
a variable star formation efficiency can produce the observed
mass-radius relation of clusters forming from GMCs. 
Part of the motivation for this approach is that, as noted in Section~3,
if GMCs {\it are}
the precursors to globular clusters, the
star formation efficiency must be much higher in GMCs in the high-pressure
ISM of starbursts than in GMCs in the disk of the Milky
Way.  The question is therefore whether a variable
star formation efficiency can wipe out the mass-radius relation of GMCs
{\it and} produce cluster systems with properties
consistent with the observed distributions of globular cluster 
masses and radii.

In the following calculations we investigate the consequences for the
mass-radius relation of clusters of a variation
in the star formation efficiency 
 with: (i) the binding energy per unit mass
of clouds; (ii) the density of clouds. In both cases, we define 
$\epsilon$ to be the star formation efficiency such that
$$
\epsilon = \frac{ M_* }{ M_c }
\eqno(4.1)    
$$
where $M_*$ is the mass of the star cluster resulting from a cloud 
mass $M_c$.

To determine the mass-radius relation of
clusters, it is clear that we also need to be able to calculate
the final cluster radius, $r_*$, produced by a cloud with radius $r_c$. 
For any star formation efficiency less than unity one expects
that $r_* > r_c$, with lower efficiencies producing larger (and possibly
unbound) clusters.  We assume that a cloud fragments and produces an initial 
cluster of radius $r_c$. If the gas loss is slow (i.e., it occurs
on timescales 
longer than the cluster dynamical time) the product of mass and radius
is an adiabatic invariant. Under these conditions, the final cluster
radius is related to the star formation efficiency by:
$$
\frac{ r_* }{ r_c } \simeq \epsilon ^{-1}.
\eqno(4.2)                               
$$
(Hills 1980; Richstone \& Potter 1982; Mathieu 1983). This expression
has recently been verified numerically by Geyer \& Burkert (2001)
for the case of slow mass loss. For more rapid mass loss, these
authors find larger expansion rates at a given $\epsilon$, but for
$\epsilon < 0.4$ the clusters are unbound.

While the dependence of cluster expansion on star formation efficiency
seems well established, there are a few
complications in directly comparing the radii of clusters
with the radii of their precursor clouds.
We have assumed above that the precursor cloud fragments
to a cluster which initially has the same radius as the cloud, $r_c$.
However, the density profiles of GMCs and globular clusters have
a different slope and form, with GMCs being described by shallow 
power-laws and most globular clusters being well-fit by a steeper King profile.
There is growing evidence that {\it young}
star clusters have power-law density profiles, but they
are still steeper than those of GMCs. For example, density profiles
inside GMCs are typically around $R^{-2}$ to $R^{-1}$, with clumps inside
clouds having even shallower profiles (Elmegreen \&
Falgarone 1996). In young LMC clusters there is a range of power-law
slopes with $R^{-2.5}$ being typical (Elson et al 1987), while Whitmore
et al (1999) find a slope consistent with this value
for a very young (6--8~Myr) globular
cluster in the merger NGC~4038/9. This difference in slope between
the density profiles of clouds and clusters means that their must be some
evolution of the density profile, and thus the radius, before or during
the fragmentation of a cloud into a cluster. This issue is compounded
by the fact that the observed radii of clouds and globular clusters are 
determined in different ways, with the cluster radius usually being a half-light
radius. Finally, 
the details of cloud fragmentation may be important in determining
the final cluster radius. For example, it is possible that fragmentation
in the center of a GMC leads to contraction of the remaining gaseous
cloud before these regions fragment.

\subsection{Star formation efficiency variation with cloud binding energy}

The binding energy per unit mass
 of a self-gravitating system scales as $M/r$. If
we assume that the star formation efficiency varies as some power
of binding energy per unit mass of the progenitor clouds, we can write
$$
\epsilon \propto \left( \frac{ M_c }{ r_c } \right)^n
\eqno(4.3)
$$
From equation (2.3) it follows that
$$
\epsilon \propto M_c^{n/2}P_s^{n/4}
\eqno(4.4)
$$
and thus
$$
M_* \propto M_c^{(n+2)/2}P_s^{n/4}
\eqno(4.5)
$$
Equations (3.3) and (4.2) yield:
$$
r_* \propto r_c^{(1-n)}P_s^{-n/2}
\eqno(4.6)
$$
Equations (4.5) and (4.6) can be used, along with equation (3.3), to
obtain a mass-radius relation for the resulting clusters:
$$
r_* \propto M_*^{(1-n)/(n+2)}
            P_s^{[1-2(n+1)]/[2(n+2)]}
\eqno(4.7)
$$
Note that for $n=0$ these expressions give $M_*\propto M_c$,
$r_*\propto r_c$, and $r_* \propto M_*^{1/2}P_s^{-1/4}$ as expected.

The weak or absent mass-radius correlation of globular clusters is
reproduced if the exponent on $M_*$ in (4.7) is close
to zero.  This occurs when $n \simeq 1$. For the specific case of
the young globular clusters in NGC~3256, Zepf et al (1999) found
$r_* \propto M_*^{0.1 \pm 0.1}$ (assuming a constant cluster mass-to-light ratio)
which is reproduced by $n \simeq 0.75 \pm 0.25$.
The value $n=1$ is interesting since it
corresponds to the case of a star formation efficiency which is directly
proportional to the binding energy per unit mass of the precursor
gas clouds (see also Section~5 below). An increase in star formation
efficiency with velocity dispersion (which scales as the square root
of binding energy per unit mass) has been suggested by Elmegreen et al
(1993) and Elmegreen \& Efremov (1997).

Equation (4.5) also allows us to investigate the variation in star
formation efficiency, and thus the upper mass of clusters, with pressure.
This is important because, as noted in Section~1, we require a much
higher star formation efficiency in GMCs in the high-pressure ISM of
starbursts than occurs in GMCs in the Milky Way. Since the above analysis
only provides scaling relations, we need to normalize the relations to
investigate this question quantitatively. We suppose
that $\epsilon$ reaches a maximum value of 0.5 for $M_c=10^7 M_{\odot}$
and $P_s = 10^8k$~cm$^{-3}$K. The values are somewhat arbitrary, corresponding
to a characteristic cloud surface pressure
in the ISM of starbursts and a cloud mass sufficient to produce a high-mass
globular cluster.
Thus, for $n=1$, (4.5) becomes:
$$
\frac{ M_* }{ M_c } = 0.5 \left( \frac{ M_c }{ 10^7 M_{\odot} }\right)^{1/2}
                     \left( \frac{ P_s }{ 10^8 k~{\rm cm}^{-3}{\rm K} }
			\right)^{1/4}
\eqno(4.8)
$$
Thus the efficiency of star formation in the Milky Way ISM with
$P_s = 10^5 k~{\rm cm}^{-3}{\rm K}$ is around 0.1 for $M_c=10^7M_{\odot}$
and $0.01$ for $M_c=10^5M_{\odot}$. While these efficiencies are
somewhat higher than observed, they do suggest that the formation
of massive, bound star clusters is much less likely in the ISM of the
Milky Way than in the ISM of starbursts.

\subsubsection{Constraints on the scenario}

Important constraints on the dependence of star formation efficiency on
cloud properties are provided by observations of the cloud and cluster
masses and mass distributions. We relate the mass distributions of
clouds and clusters by
$$
N(M_*)dM_* \propto N(M_c) \left( \frac{ dM_c }{ dM_* } \right) dM_*
\eqno(4.9)
$$
A variable star formation efficiency will, in general, lead to different
slopes for the cloud and cluster mass distributions.
Equations (4.5) and (4.9) can be used to quantify the difference in slope:
$$
\alpha = \frac{ 2 \beta + n }{ n + 2 } .
\eqno(4.10)
$$                  
(For $n=0$, $\alpha=\beta$ as expected.) Alternatively,
$$
n = \frac{ 2(\beta - \alpha) }{ \alpha -1 }
\eqno(4.11)
$$

Equation (4.11) allows us to use the observed values of $\alpha$ and
$\beta$ to constrain the permitted range of $n$. Since $\alpha$ and
$\beta$ are comparable, it is apparent that small values of $n$
are favored.  For example, for the mass function slopes given in
Section~2 of $\beta=2$ and $\alpha=1.8$ one obtains $n=0.5$, giving
a mass-radiius relation consistent with that of the young clusters
in NGC~3256.
The observational uncertainty on $\alpha$
and $\beta$ means that $n=1$ is marginally consistent with the data.

Note that if a variable star formation efficiency of this form is responsible
for wiping out the cloud mass-radius relation, it is inevitable
that the cluster mass function is shallower than the precursor
cloud mass function. In other words, our picture predicts that
$\alpha < \beta$.  Thus determinations of the distributions of cloud and 
cluster mass have the potential to
refute or support this scenario. Unfortunately, the current uncertainties
in these quantities, as well as the fact that cluster and cloud mass
spectra are rarely derived for the same systems, do not allow a definitive
test of the scenario.
However, future observations of the mass functions
of molecular clouds and globular clusters may be able to strongly
constrain the dependence of star formation efficiency on binding
energy.

\subsection{Star formation efficiency variation with cloud density}

Another case of potential interest is a variation of star
formation efficiency with cloud density.
If we suppose that $\epsilon$ has a power-law dependence on density, we
can write:
$$
\epsilon \propto \rho ^m
\propto \left( \frac{ M_c }{ r_c^3 } \right) ^m 
\eqno(4.12)
$$
Following the same treatment as in Section~4.1 we obtain the following
relations:
$$
\epsilon \propto M_c^{-m/2}P_s^{3m/4}
\eqno(4.13)
$$
$$
M_* \propto M_c^{(2-m)/2}P_s^{3m/4}
\eqno(4.14)
$$ 
$$
r_* \propto r_c^{(1+m)}P_s^{-m/2}
\eqno(4.15)
$$    
$$
r_* \propto M_*^{(1+m)/(2-m)}
            P_s^{[3-(m+1)4]/[2(2-m)]}
\eqno(4.16)
$$    
Again, the behavior of these expressions when $m=0$ is as expected.

The exponent on $M_*$ in this relation
can not equal zero for positive $m$. Thus even though relationships
like the Schmidt law ($\epsilon \propto \rho ^2$) provide a good
description of global star formation efficiency, such a relationship
can not wipe out the mass-radius relation of star clusters forming
from GMCs, nor any other system of pressure-supported
progenitor clouds obeying the virial
theorem. The failure of the widely applied Schmidt law and similar
relationships between star formation efficiency and density to
account for the observation that cluster mass and radius are
independent demonstrates the importance of this observation
for constraining globular cluster formation models.
  
\subsection{Pressure variations in the ISM}
 
The above discussion of the effects of a variable star formation efficiency
on the mass-radius relation of globular clusters implicitly assumes a constant
ISM pressure. Thus for a given cloud mass, the cloud radius and resulting
cluster radius is uniquely determined. In practice, it seems inevitable that
the ISM pressure in starbursts exhibits some variation with location.
One immediate consequence of including such pressure variations is that the
precursor GMCs occupy a band rather than a line in the mass-radius
plane. In the context of our scenario, this means that clouds at a given
mass will produce globular clusters with a range of densities.
A large range in density at a given mass is a notable feature
of the globular cluster system of the Milky Way. The same broad
distribution of densities appears to be the case for young globular 
clusters in NGC~3256. It is currently difficult to 
link quantitatively the pressure variations in the
ISM and the observed density range of young and old globular cluster systems. 
Old globular clusters have likely undergone 
significant mass-loss, while the densities of young globular
clusters are not well determined. 
Future observations of
young globular cluster systems and the pressure in the surrounding ISM
may produce useful constraints on the relation between these
quantities.

An additional issue is whether pressure variations and the resulting
dispersion in the GMC mass-radius relation have any impact on the
mass-radius relation of the resulting clusters derived above.
We have carried out a preliminary study of this question using
Monte Carlo simulations that generate an ensemble of clouds
with a mass distribution like that observed for GMCs. Each cloud
is randomly assigned a pressure drawn from a log-normal distribution
with a peak value consistent with that of the ISM in starbursts. 
The dispersion of this log-normal distribution was varied with
simulations for one, two and three dex performed.
Individual cloud radii are calculated from the assigned mass and
pressure of each cloud using the Ebert-Bonner relations (3.1) and
(3.2). The final globular cluster masses and radii are obtained
by using one of the prescriptions for star formation efficiency
described above.
 
We have found that pressure variations alone can {\it not} wipe out the virial
mass-radius relation of the precursor clouds. That is, if we assume
a constant star formation efficiency, the resulting clusters still
exhibit the scaling relation $r \propto M^{0.5}$ of the GMCs.
We have also used this technique to study the effects of pressure
variations for the case of a star formation efficiency that
varies linearly with cloud binding energy per unit mass
(i.e., the case $n=1$, see Section~4.1).  In this case, the dispersion in the 
mass-radius relation produced by pressure variations has the effect
of slightly weakening constraints derived for the constant pressure
situation. This is simply because the dispersion in the mass-radius
relation produces an uncertainty in the measured slope of the mass-radius
relation, at least for a finite sample size. To quantify this somewhat,
recall that in the constant pressure case a relation of the form 
$r_* \propto M_* ^y$ is produced with
$y$ being dependent on $n$. Our simulations suggest
that pressure variations of the order described above will produce cluster
masses and radii that are indistinguishable from no mass-radius
relation if $y$ is less than about 0.1, given a sample size
comparable to the number of young clusters observed in NGC~3256. 

\section{Discussion}

Observational advances over the last decade mean that theories of
globular cluster formation can now be motivated and constrained by
observation. We have argued in this paper that the observed formation
of globular clusters in starbursts and mergers reveals the conditions
under which globular cluster formation occurs.
We have shown that the high-pressure of
the ISM of mergers and starbursts implies that GMCs in this environment
have the right properties to be globular cluster precursors, if they
form stars with high efficiency. A high star formation efficiency in these
dense GMCs may be a natural consequence of their short dynamical
timescales and high binding energies. 
However, all GMCs studied to date follow
a strong scaling relation between mass and radius, whereas the correlation
between mass and radius for globular clusters appears to be 
weak or non-existent.
 We have suggested that a star formation
efficiency that is a function of cloud binding energy can wipe out the
mass-radius relation of the progenitor GMCs. Such a variation automatically
implies a much higher star formation efficiency in GMCs in the high-pressure
environment of starbursts than in the low-pressure ISM of ordinary disk
galaxies like the Milky Way. 

There are plausible reasons why star formation efficiency might
be roughly proportional to the binding energy per unit mass of clouds
(see also Elemegreen et al 1993; Elmegreen and Efremov 1997). 
To a first approximation, the disruptive energy input from massive
stars will be proportional to the number of such stars and thus the 
mass of the cloud, hence the normalization of binding energy to unit mass.
While the details of the disruption produced by massive stars is beyond
the scope of this paper, it seems likely that clouds with a higher
binding energy will be less affected by such disruption and therefore
convert a higher fraction of their gas mass into stars. 
We plan to address this question with simulations in a future paper.

A key ingredient of our scenario stems from the observation that the mass and radius
distributions of GMCs and globular clusters are
similar, whereas their mass-radius relations are not. This observation
constitutes a challenge for any scenario in which 
GMCs (or other virialized gas clouds) are globular cluster progenitors.
Specifically, it is difficult to eliminate the original mass-radius relation
without significantly altering either the mass distribution or
the radius distribution. We have shown that a star formation efficiency that
varies with cloud binding energy leads to the testable prediction that
the globular cluster mass distribution is shallower than that of
the clouds. Constraints based on the radius distribution of clouds and
clusters are more difficult
to apply. This is partly because the radius distribution of GMCs is a simple
power-law whereas that of young clusters is not.
Thus it is not possible to directly compare the slopes of cloud
and young cluster power-laws and obtain constraints on the
scenario as we did for the mass distributions. Other complicating
factors include the difference
in the density profiles of clouds and clusters (suggesting the radius
changes as the density profile steepens from the cloud to cluster slope)
 and the fact that cloud and cluster
radii are determined in different ways. 

We note that the expansion of clusters due to adiabatic mass
loss has been exploited by other authors in an attempt to
understand the properties of old globular clusters
(e.g.\ Gunn 1980; Djorgovski 1991; Djorgovski \& Meylan 1994). 
However, these previous studies assumed a different mass-radius
relation for the ``initial'' clusters. More importantly,
in the present scenario the ``lost'' 
mass is simply gas mass that is not converted into stars. 
The studies mentioned above assume stellar mass-loss is responsible for
cluster expansion and an evolution of the mass-radius relationship.
The observation that young cluster systems do not 
follow the mass-radius relation of the precursor clouds
tends to favor early loss of
mass from the clusters, as envisioned in our scenario. 

While our picture addresses the formation of globular clusters at the
current epoch in mergers and starbursts, it does not directly address
the formation of old globular clusters found around galaxies from
dwarfs to massive spirals and ellipticals. We have previously argued
(Ashman and Zepf 1992) that many of the old globular clusters around
ellipticals formed in mergers at earlier epochs. As such, the formation
of these globular clusters is directly analogous to those forming in mergers
at the current epoch and addressed in this paper. In a future paper we
will investigate whether the physical conditions that give
rise to globular clusters in starbursts and mergers are present
in young dwarf galaxies and spirals.

\acknowledgments

We are grateful to useful discussions with Dean McLaughlin and 
Richard Larson. This work was supported in part by NASA Astrophysics
Theory Grant NAG5-9168.

\end{document}